\documentclass[12pt]{book}

\usepackage[dvips]{graphicx,color}
\usepackage{makeidx,tsukuba}

\makeauthorindex
\makeindex

\begin{document}

\BookTitle{\itshape The 28th International Cosmic Ray Conference}
\CopyRight{\copyright 2003 by Universal Academy Press, Inc.}
%\tableofcontents
\pagenumbering{arabic}

\chapter{Intensive TeV Gamma-Ray and X-Ray Observations of the Blazar Mrk 421 in December 2002 and January 2003}

\author{
P.F. Rebillot,$^{1,2}$ , S.B. Hughes$^1$, I.H.~Bond, P.J.~Boyle, S.M.~Bradbury, J.H.~Buckley, D.~Carter-Lewis, O.~Celik, W.~Cui, M.~Daniel, M.~D'Vali, I.de~la~Calle~Perez, C.~Duke, A.~Falcone, D.J.~Fegan, S.J.~Fegan, J.P.~Finley, L.F.~Fortson, J.~Gaidos, S.~Gammell, K.~Gibbs, G.H.~Gillanders,  J.~Grube, J.~Hall, T.A.~Hall, D.~Hanna, A.M.~Hillas, J.~Holder, D.~Horan, A.~Jarvis, M.~Jordan, G.E.~Kenny, M.~Kertzman, D.~Kieda, J.~Kildea, J.~Knapp, K.~Kosack, H.~Krawczynski, F.~Krennrich, M.J.~Lang, S.~LeBohec, E.~Linton, J.~Lloyd-Evans, A.~Milovanovic, P.~Moriarty, D.~Muller, T.~Nagai, S.~Nolan, R.A.~Ong, R.~Pallassini, D.~Petry, B.~Power-Mooney, J.~Quinn, M.~Quinn, K.~Ragan, P.T.~Reynolds,  H.J.~Rose, M.~Schroedter, G.~Sembroski, S.P.~Swordy, A.~Syson, V.V.~Vassiliev, S.P.~Wakely, G.~Walker, T.C.~Weekes, J.~Zweerink \\
{\it (1) Department of Physics, Washington University, St.Louis MO 63130, USA}\\
{\it (2) The VERITAS Collaboration--see S.P.Wakely's paper} ``The VERITAS Prototype'' {\it from these proceedings for affiliations}
}

\section*{Abstract}
We report on observations of Markarian 421 made with the Whipple 10m \v{C}erenkov telescope and the RXTE satellite during a multi-wavelength campaign in December 2002 and January 2003, initiated by a Whipple target of opportunity.  The observations revealed several flares with flux levels between 1 and 2 times the flux of the Crab Nebula.  We will discuss the temporal properties, including evidence of X-ray/TeV gamma-ray flux correlation.  

\section{Introduction}

The BL Lac object Markarian 421 (Mrk 421) is a TeV blazar at redshift z=0.031. Along with the other TeV blazars such as Markarian 501 and 1ES1959+650, it has been the focus of extensive multi-wavelength observations in the X-ray and gamma-ray regimes.  In the past, Mrk 421 observations have had insufficient time sampling \cite{buckley}, coupled with short variability timescales.  This has made it difficult to draw conclusions concerning the correlation between fluxes seen in the X-ray and gamma-ray regimes.  Recent multi-wavelength campaigns have attempted to correct this, and have shown complex multi-wavelength behavior \cite{jordan},\cite{takahashi}.  For this reason, a target of opportunity (ToO) multi-wavelength campaign with the PCA instrument aboard the RXTE satellite was proposed.  In November 2002, Mrk 421 was observed with the Whipple 10m telescope, and was seen at a level of 3 times the flux of the Crab Nebula.  These observations triggered the X-ray observations and multi-wavelength campaign during December 2002 and January 2003.  Insights gained from any correlation between X-ray and gamma-ray fluxes will help determine the source of these high energy photons, be it from inverse Compton scattering of synchrotron photons off the same electrons, or from some other model, including various non-leptonic models.

\section{Data}
\subsection{Whipple 10m Data}
Mrk 421 was observed with the Whipple 10m telescope during the period December 4 2002 UT and January 15 2003 UT.  26 ON/OFF mode runs were taken,  resulting in 12 hours of data.  49 runs were taken in TRACKING mode, resulting in 20 hours of data.  For this analysis,  all ON-source data is analyzed with the tracking analysis, resulting in a total exposure of 32 hours.  Details about the present configuration of the Whipple telescope, including the GRANITE-III camera, are given elsewhere \cite{finley}.  The data was then analyzed using a standard moment-parameterization technique, $\verb SuperCuts 2000$.  In tracking analysis,  parameterized \v{C}erenkov showers initiated by gamma rays that do not point back to the source are treated as background.  The TRACKING ratio, which is defined as the ratio of the number of background events to the number of ON-source events,  is calculated from the 26 OFF mode runs taken for this source.

\subsection{RXTE PCA Data}
The X-ray data described in the following is based on the 3-25~keV data from the Proportional Counter Array \cite{jahoda}  on board the {\it RXTE} satellite.  We did not use the 15-250 keV data from the High-Energy X-ray Timing Experiment \cite{rothschild} due to poor signal-to-noise ratio. Standard-2 mode PCA data gathered with the top layer of the operational PCUs (Proportional Counter Units) were analyzed.  The number of PCUs operational during a pointing varied between 2 and 4.

After applying the standard screening criteria and removing by hand abnormal data spikes, the net exposure in each Good Time Interval ranged from 168 secs to 9.01 ksecs. Lightcurves were then extracted with \verb+FTOOLSv+5.0. Background models were generated with the tool \verb+pcabackest+, based on the {\it RXTE} GOF calibration files for a ``bright''source (more than 40 counts/sec/PCU). Response matrices for the PCA data were created with the script \verb+pcarsp+v.7.11.

\section{Results}

Figure 1 gives the  X-ray and gamma-ray light-curves for this Dec 2002 to Jan 2003 multi-wavelength campaign.   Analyzing the 26 OFF runs in the gamma-ray data result in a TRACKING ratio of $\alpha = 3.786$.  The gamma-ray error bars are calculated using an approximation to the Li and Ma method \cite{lima}.  Corrections for the varying zenith angle during the observation of Mrk 421 are not taken into account here.

\begin{figure}[t]
     \begin{center}
       \hspace{-0.8cm}
       \includegraphics[height=4.2in]{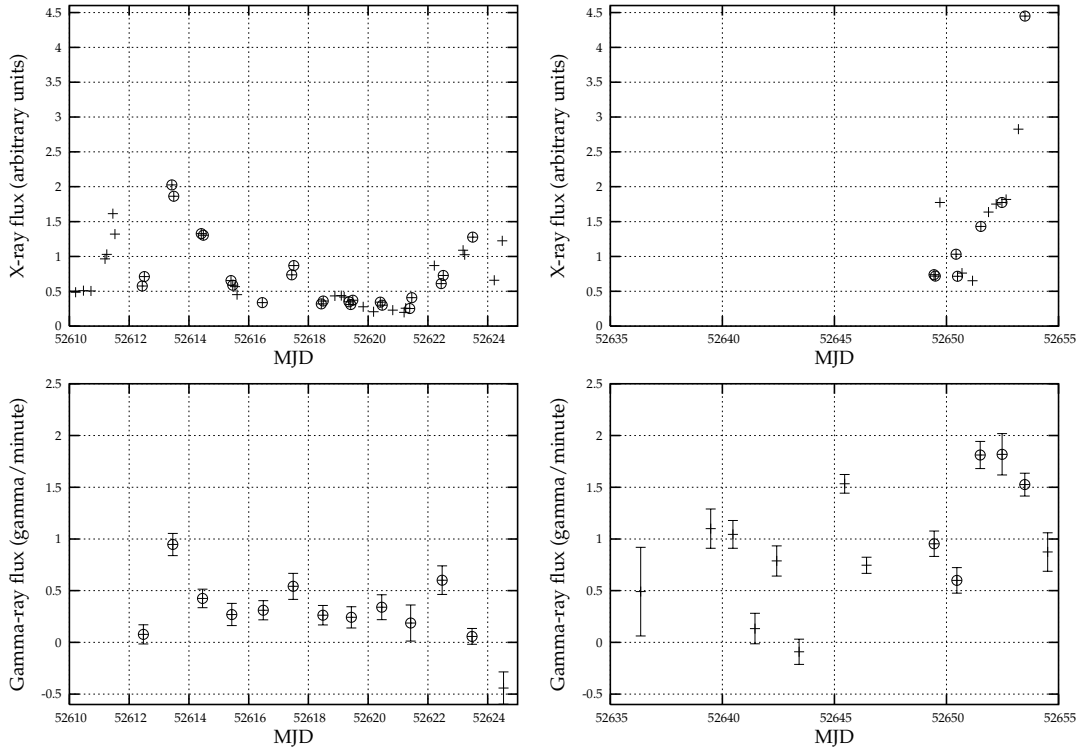}
     \end{center}
  \vspace{-2.5pc}
  {\small
  \caption{Data from RXTE satellite and Whipple 10m telescope taken during Dec 2002 and Jan 2003, over 45 days.  The Whipple data shown here are per night averages.  For clarity, the error bars on the RXTE data have not been included, since they are smaller than the size of the data points.  Data points that are circled correspond to periods when gamma-ray and X-ray observations were simultaneous (within 5 minutes).}}
\end{figure}

\section{Discussion}

By visual inspection of the data in Figure 1 from the first half of the multi-wavelength campaign (MJD 52610 - 52625),  it is possible that a correlation exists between the X-ray and gamma-ray light-curves.  In particular, the small flare between MJD 52613 and 52614 peak at the same time in both the X-ray and gamma-ray regimes.  Figure 2 gives the discrete correlation function \cite{edelson} of the combined data set.

\begin{figure}[t]
     \begin{center}
         \includegraphics[height=2.0in]{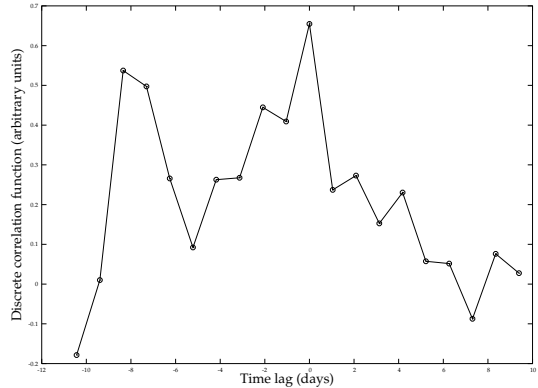}
     \end{center}
  \vspace{-1.5pc}
  {\small
  \caption{Discrete correlation function of the complete X-ray and gamma-ray data set.  A positive time lag means the gamma-ray flux precedes the X-ray flux.}}
\end{figure}

\section{Conclusion}

With this past multi-wavelength campaign, it appears that there may be a complicated correlation between the X-ray and gamma-ray fluxes for Mrk 421.  By considering the seemingly complex structure of these flares, it is hoped that clues into the production mechanism can be found.  Continued multi-wavelength observations of Mrk421 are warranted, considering interesting recent observations \cite{krawcz} of 1ES1959+650 showing possible orphan flares.\footnote{We acknowledge the technical assistance of E.Roache and J. Melnick.  This research is supported by grants from the U.S. Department of Energy, by Enterprise Ireland and by PPARC in the UK.}

\endofpaper
\end{document}